\newcommand{\PRB}[3]{Phys.\ Rev.\ B\ {\bf #1},\ #2 (#3)}

\newcommand{\PRX}[3]{Phys.\ Rev.\ X\ {\bf #1},\ #2 (#3)}

\newcommand{\NATP}[3]{Nat. Phys.\ {\bf #1},\ #2 (#3)}

\newcommand{\appsection}[1]{\let\oldthesection\thesection
  \renewcommand{\thesection}{ \oldthesection}
  \section{#1}\let\thesection\oldthesection}
\documentclass[twocolumn,preprintnumbers,amsmath,amsfonts,amssymb,notitlepage,showpacs,prl]{revtex4-1}
\usepackage{geometry}
\date{\today}
\DeclareMathAlphabet{\mathpzc}{OT1}{pzc}{m}{it}
\usepackage[utf8x]{inputenc}
\usepackage{amsmath}
\usepackage{amsfonts}
\usepackage{amssymb}
\usepackage{graphicx}
\usepackage{nicefrac}
\usepackage{amsbsy}
\usepackage{float}
\usepackage{bm}
\usepackage{epstopdf}
\usepackage{dsfont}
\usepackage{siunitx}
\usepackage{subfigure}
\usepackage{color}

\def \be{\begin{equation}}
\def \ee{\end{equation}}
\def \ba{\begin{array}}
\def \ea{\end{array}}
\def \bea{\begin{eqnarray}}
\def \eea{\end{eqnarray}}

\begin{document}
\title{{Frustration induced quasi-many-body localization without disorder}}
\author{Sayan Choudhury,$^{1}$ Eun-Ah Kim,$^{2}$ and Qi Zhou$^{1}$}
\email{zhou753@purdue.edu}
\affiliation{$^{1}$Department of Physics and Astronomy, Purdue University, 525 Northwestern Avenue, West Lafayette, IN 47907, USA\\
$^{2}$Department of Physics, Cornell University, Ithaca, New York 14853, USA}
\begin{abstract}
Motivated by the question of whether disorder is 
a prerequisite for localization to occur 
in quantum many-body systems, 
we study a frustrated one-dimensional spin chain, which supports localized many-body eigenstates in the absence of disorder. When the system is prepared in an initial state with one domain wall, it exhibits characteristic signatures of  quasi-many-body localization (quasi-MBL), including initial state memory retention,  an exponentially increasing lifetime with enlarging the size of the system, a logarithmic growth of entanglement entropy, and a logarithmic light cone of an out-of-time-ordered correlator. We further show that the localized many-body eigenstates can be manipulated as pseudospin-1/2s and thus could potentially serve as qubits. Our findings suggest a new route of using frustration to access quasi-MBL and preserve quantum coherence. 
\end{abstract}
\maketitle
The localization of a single quantum particle due to disorder has been well understood by Anderson localization \cite{Anderson1958}. In recent years, the interacting version of Anderson localization - many-body localization (MBL)- has received a lot of attention\cite{BLL2006,PalHuse2010,MoorePRL2012,HusePRB2014MBL,KhemaniMBLPRX,VoskMBLPRX,SerbynPRLMBL,AltlandPRLMBL2017,
HuseMBLPRB2013,NayakTC,YaoTC,BahriNatComm,YaoQCMBL1,YaoQCMBL2,NandkishoreMBLrev,BlochMBL1,BlochMBL2,MonroeMBL}. 
Many-body localized systems fail to thermalize under their own dynamics, leading to a breakdown of 
the conventional paradigm of quantum statistical mechanics and several intriguing applications, ranging from realizing phases of matter forbidden in equilibrium \cite{HuseMBLPRB2013,NayakTC,YaoTC,BahriNatComm} to quantum information processing \cite{YaoQCMBL1,YaoQCMBL2}. Recent experiments with ultracold atoms \cite{BlochMBL1,BlochMBL2} and trapped ions 
have demonstrated the existence of the many-body localized phase \cite{MonroeMBL}, as well as time-crystals, a new state of matter brought by MBL in periodically driven systems \cite{MonroeTC}. 

Whereas early studies of MBL required the presence of static disorder, some recent efforts 
have attempted to address whether essential aspects of MBL can be observed in disorder free systems. Several of these proposals require multiple species or degree of freedoms. For instance, 
some researchers have put forward a general class of models involving two species of particles, one heavy and one light. 
The heavy particles produce a quasi-static disorder potential for the light particles, leading to localization \cite{YaocleanMBL,GrovercleanMBL,MullercleanMBL,AbanincleanMBL}. It has been suggested that this class of translation-invariant systems could exhibit quasi-MBL, where characteristic features of MBL can be observed up to certain time scales, and ergodicity is restored at longer times. Other  works have exploited quantum versions of classical glass models \cite{GarrahanJSM,GarrahanPRB} and models involving both spin and fermionic degrees of freedom, the latter of which have interesting connections to lattice gauge theories \cite{SmithPRLMBL,ScardicchioPRLMBL2018}.

In this Letter, we take a different route and show that quasi-MBL can be accessed in a frustrated 1D spin chain, which offers a new platform for preserving quantum memories in many-body systems. Compared to other systems supporting MBL and quasi-MBL, this system has a number of merits in both fundamental and practical aspects. Unlike systems with disorder or quasi-periodic potentials, all single-particle states in our systems are extended ones in real space. The localized states originate purely from many-body effects, rising from the interplay between interaction and frustration. Different from multi-species models where multiple energy scales are important, here, the main ingredient in determining the lifetime of the quantum memory is the size of the system. In experiments, it is relatively easy to realize such a vanilla model, as it requires only one species of particles. Our results thus suggest that frustration could serve as a new breeding ground for studying quasi-MBL in both theory and experiments.

\begin{figure}
\includegraphics[scale=0.3]{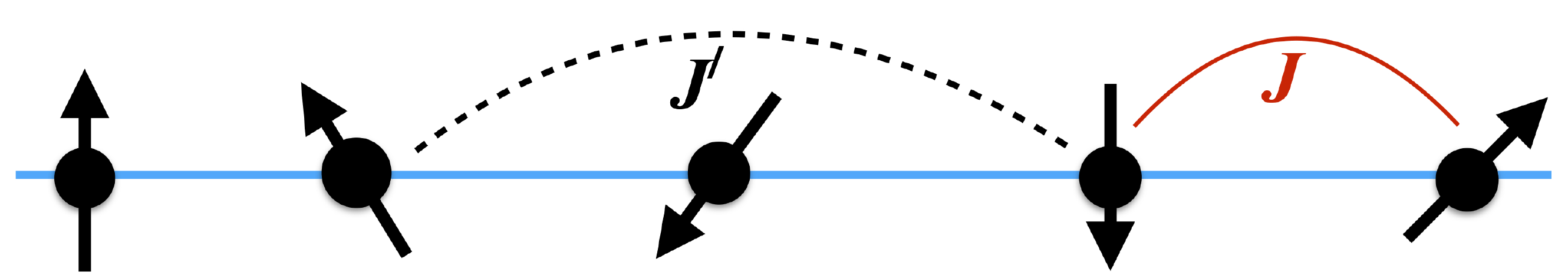}
\caption{ Schematic picture of the frustrated one-dimensional spin chain given in eq.(1). For a wide range of values of $J^{\prime}/J_z$, this model exhibits signatures of quasi-MBL.}
\label{fig0}
\end{figure}

As shown in Fig.\ref{fig0}, our model contains both the nearest neighbor coupling $J$ and the next nearest neighbor coupling $J'$. The Hamiltonian is written as
\bea
H = \sum_i J_z {S}^z_i{S}^z_{i+1}+(\frac{J}{2} S_i^{+}S_{i+1}^{-}+\frac{J^{\prime}}{2} S_i^{+}S_{i+2}^{-}+h.c), \nonumber\\
\label{H}
\eea
where $S^\nu_i = \frac{\hbar}{2} \sigma^{\nu}_i$, $S^{\pm}_i = S^x_i \pm i S^y_i $, $\sigma^{\nu=x,y,z}_i$ being a Pauli matrix defined at site $i$, and $J^{\prime}/{J_z}<0$. 

\begin{figure}
\includegraphics[scale=0.35]{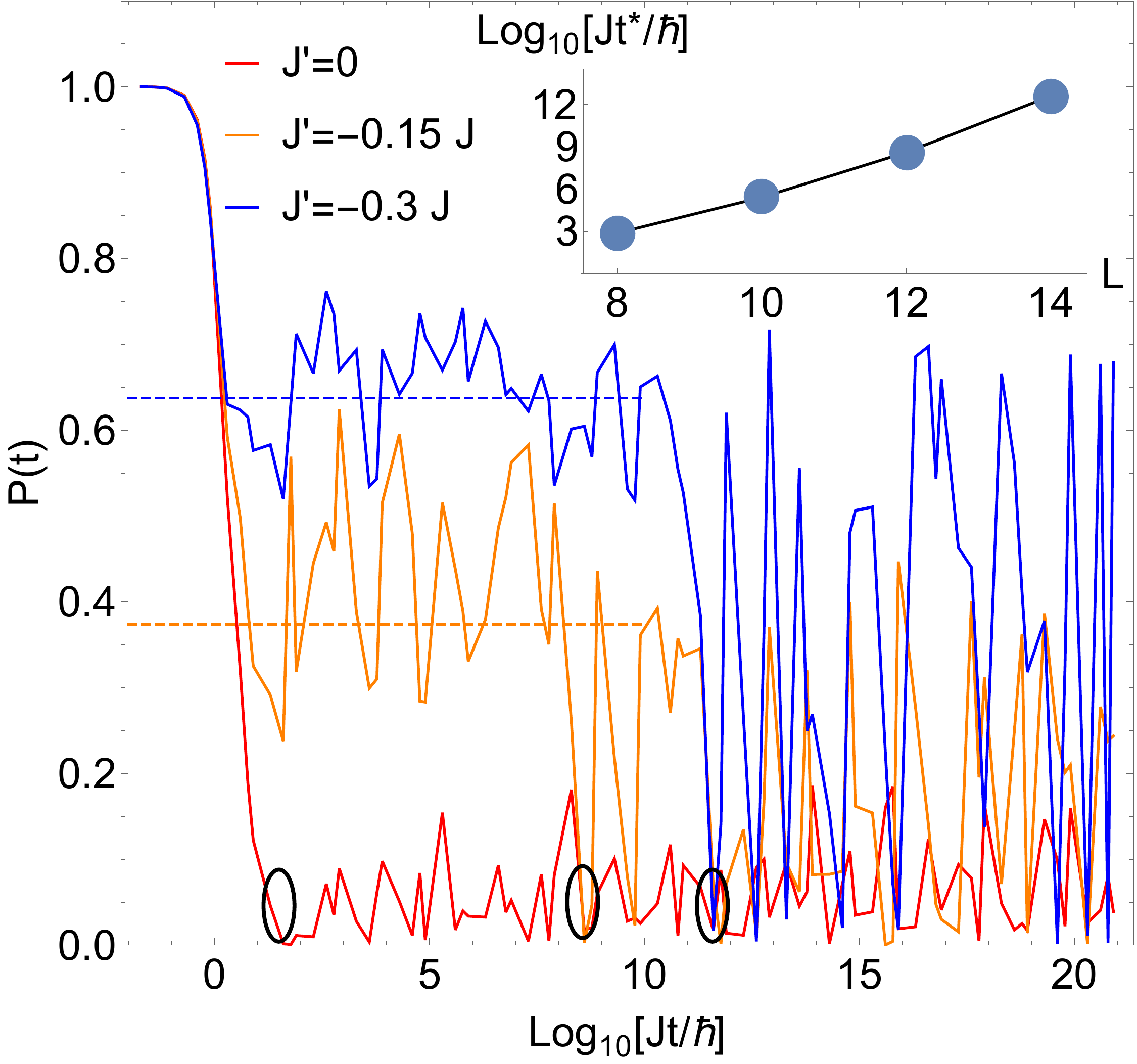}
\caption{
The memory retention, $P(t) = \vert \langle \psi(0)\vert\psi(t) \rangle \vert^2$, of an initial state with one domain wall ($\vert \uparrow \ldots \uparrow  \downarrow \ldots \downarrow \rangle$), when $J_z= J$. For a wide range of $J^{\prime}/J_z$, $P(t)$ remains finite for a long time. Dashed horizontal lines represent $|\alpha|^4$ (see text). The black circle denotes the quasi-MBL lifetime, $t^{*}$. The inset shows $t^{*}$ (when $J^{\prime} = -0.15 J_z$), which increases exponentially with the system size.}
\label{fig1}
\end{figure}

Whereas the ground states of such frustrated models and related ones have been explored
thoroughly in the literature \cite{Arconnatcomm,LauchliPRB2009,FurusakiPRB2008}, quantum dynamics has been less studied. Here, using exact diagonalization and an effective model, we solve the Hamiltonian in Eq. (\ref{H}) in a lattice with $L$ sites and open boundary conditions. We show that this Hamiltonian gives rise to  essential signatures of quasi-MBL, including initial state memory retention, an exponentially increasing lifetime with enlarging
the size of the system, a logarithmic growth of entanglement entropy, and a logarithmic light cone of an out-of-time-ordered correlator. Our results suggest that a large class of models with frustration could serve as a new breeding ground for quasi-MBL, an aspect not addressed before.

{\it Initial State Memory Retention:} 
A direct test of MBL and quasi-MBL is
examining the initial state memory retention, 
i.e., the overlap of the state at a later time $t>0$
with the initial one at $t=0$:
\be
P(t) = \vert \langle \psi(0)\vert\psi(t) \rangle \vert^2,
\label{le}
\ee
which is a type of  Loschmidt echo. 
The rationale behind this diagnostic is that the memory of the initial state decays exponentially in ergodic systems, while for MBL and quasi-MBL phases, $P(t)$ is expected to remain finite as time goes on. In quasi-MBL, the memory of the initial quantum state is retained only up to a certain time scale, which grows exponentially with increasing the size of the system, as shown later.

We consider an initial state $\vert\psi_I\rangle=|\uparrow \ldots \uparrow \downarrow \ldots \downarrow\rangle$ (or its mirror image $\mathcal{I}|\psi_I\rangle=\vert\downarrow \ldots \downarrow \uparrow \ldots \uparrow \rangle$), which has only one domain wall separating the left $L/2$ sites occupied by spin-up and the other half sites occupied by spin-down, as analogous to the state studied a recent ultracold 
atom experiment in two dimensions \cite{BlochMBL2}. $\mathcal{I}$ is the inversion operator that swaps the spins between the left and the right half of the chain. To simplify notations, we have chosen an even $L$ and put the domain wall at the center of the system, as denoted by o in Fig.2. When the size of the system is large enough, there is essentially no difference between $L$ and $L+1$ sites. The exact initial position of the domain wall does not lead to any qualitative difference either, if it is far away from the boundaries. 

Our numerical results are illustrated in figure 2. When $J'=0$, i.e., frustration is absent, $P(t)$ decays down to zero very fast, as expected in a disorder free system. When $J'$ is finite and $J'/J_z<0$, following an initial decay, $P(t)$ remains finite for a long time scale, $t^*$,
which is orders of magnitude larger than the single particle tunneling time scale $\hbar/J$. For instance, for $J'=-0.3J_z$, $L=12$, $P(t)$ oscillates around $0.65$ up to $t^*\approx 10^{12} \hbar/J$. These results unambiguously demonstrate that our frustrated model gives rise to a long time memory of the initial state, a characteristic feature of quasi-MBL.  

\begin{figure*}
\subfigure{\raisebox{0.0cm}{\scalebox{0.25}{\includegraphics{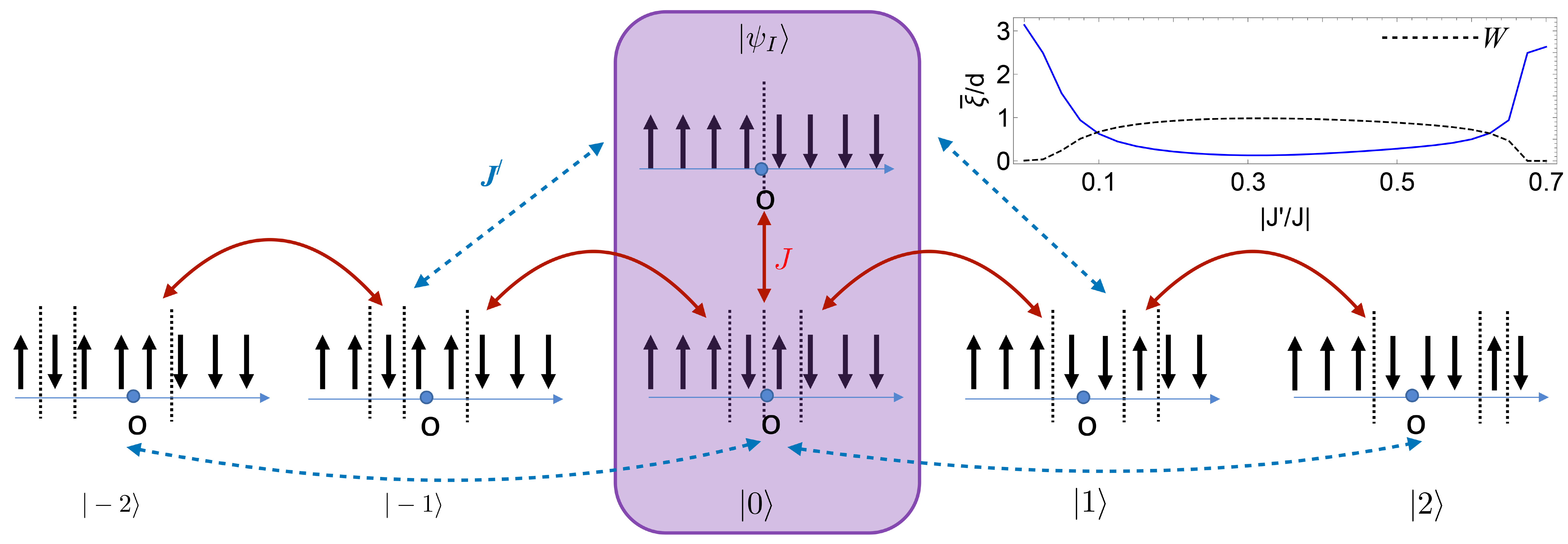}}}}\label{f2}
\caption{
Schematic of the effective model $H_{\text{eff}}$. Dashed lines represent domain walls, and o denotes the origin of the lattice. 
The inset shows the properties of the localized eigenstate: the blue(thick) line represents the localization length $\overline{\xi}$
and the black (dashed) line represents the total weight, $W$, of $|\psi_I\rangle$ and $|0\rangle$ in the eigenstate. 
Both measures show that $\vert \psi_0\rangle$ is maximally localized when $J^{\prime} \approx -0.3 J$, when $J_z=J$.}
\end{figure*}

{\it Localized eigenstates:} The underlying mechanism of the quasi-MBL can be traced to 
localized many-body eigenstates. We find that, for a large enough system, the Hamiltonian in Eq.(\ref{H}) has two nearly degenerate many-body eigenstates, $\vert\psi_0 \rangle$ and $\vert\psi_1 \rangle=\mathcal{I}\vert\psi_0 \rangle$. When $L\rightarrow \infty$, these two states become degenerate and $\vert\psi_0 \rangle$ is written as 
\be
\vert\psi_0 \rangle \approx \alpha|\psi_I\rangle+\beta_0  \vert { 0}\rangle+\sum_{j\neq 0} \beta_j  \vert { j}\rangle, \label{psi0}
\ee
where $ |{ 0}\rangle \equiv  S^{+}_{{\rm o}+\frac{1}{2}}S^{-}_{{\rm o}-\frac{1}{2}}\vert \psi_I\rangle$, $\vert { j>0}\rangle \equiv S^{+}_{{\rm o}+j+\frac{1}{2}} S^{-}_{{\rm o}-\frac{1}{2}}\vert \psi_I\rangle$, and $\vert { j<0}\rangle \equiv S^{+}_{{\rm o}+\frac{1}{2}}S^{-}_{{\rm o}+j-\frac{1}{2}}\vert \psi_I\rangle$ and $o=(L+1)/2$ denotes the center of the chain. 

As shown in Fig.3, each term in Eq.(\ref{psi0}) contains only one or three domain walls. When $j$ is positive (negative), 
$(|j|+1/2)d$ denotes the distance of a spin-up (spin-down) immersed in a ``bath" of spin-down(spin-up), which causes two confined domain walls propagating towards right (left). For simplicity, we set the lattice space $d=1$. Due to a large energy gap ($\ge 4 J_z$) measured from $\vert \psi_I\rangle$,  weights of states with even more domain walls in such eigenstate are negligible,  and the remaining single domain wall is fixed at site $L/2\pm 1$. These two eigenstates are localized in the sense that $\beta_j$ decays exponentially and domain walls are confined in a small region in the real space.
We define a localization length, 
\be
\overline{\xi} =\sum_j (|j|+1/2)|\beta_j|^2.
\ee
For instance, if $J_z=J$ is chosen, $\overline{\xi}$ reaches its minimum ($\sim 0.1$ lattice spacing),  when $J^{\prime}/J \approx -0.3$, as shown in Fig. 3.
Under this situation, the localized eigenstate then is well represented by 
a psuedospin-1/2, $\vec{\tau}$, composed of $|\psi_I\rangle$, the initial state of interest, and $|0\rangle$.

Such localization can be understood from an effective model. Projecting the full Hamiltonian to the subspace composed by states with at most three domain walls,  we obtain $H_{\rm eff}=H_0+H_1$, where
\bea
H_{0}=-2J_z |\psi_I\rangle\langle \psi_I|+(J|\psi_I\rangle\langle 0|+h.c.)
\eea
is a local Hamiltonian. $H_1$ is written as 
\bea
H_1&=&(J' (|\psi_I\rangle\langle 1|+|\psi_I\rangle\langle -1|+h.c.) \nonumber\\
&+&(J\sum_{j}|j\rangle\langle j+1|+ J'\sum_{j}|j\rangle\langle j+{2}| +h.c.).
\eea
Due to the conservation of the total magnetization in the $z$ direction, other states with three domain walls are decoupled. 
We have verified that $H_{\rm eff}$ could well reproduce both the localized eigenstates and $P(t)$ of our chosen initial state \cite{suppmat}. Using $H_{\rm eff}$, 
$\beta_{j\neq 0}$ can be expressed in terms of $\alpha$ and $\beta_0$ in a closed form \cite{suppmat}. In particular, 
in the limit where $\alpha$ is much larger than $\beta_{j}$, $\beta_0 = \frac{J}{2 J_z} \alpha$ is well satisfied, 
and $\beta_j$ can be expressed in a compact form when $J^2 \ge 4 J' J_z$, 
\be
\beta_{j}= \left(A \exp(- \vert j \vert/\xi_1) + B \exp(-\vert j \vert/\xi_2)\right) \alpha, \label{loclength}
\ee
where $A$ and $B$ are constants.  
Thus, at large distance, $\beta_j$ can be well approximated by $e^{-\vert j \vert/\xi}$, where $\xi=max\{\xi_1, \xi_2\}$. 
We can gain further insight from examining $\beta_j$, when $j$ is small,
\bea
\beta_1 &=& \frac{J}{2 J_z} \beta_0 + \frac{J^{\prime}}{2 J_z} \alpha \approx \left((\frac{J}{2 J_z})^2 + \frac{J^{\prime}}{2 J_z}\right) \alpha \\
\beta_2 &=& \frac{J}{2 J_z} \beta_1 + \frac{J^{\prime}}{2 J_z} \beta_0 \approx \left((\frac{J}{2 J_z})^3 +\frac{J J^{\prime}}{2 J_z^2}\right) \alpha
\eea
When $J^2/2J^{\prime}J_z < 0$, 
$\beta_{1,2}$, as well as other $\beta_{j\neq 1, 2}$,  
are suppressed due to destructive interference of multiple paths that a spin-up (spin-down) could take to tunnel to a site away from the center of the system. This leads to the localization of the eigenstate $\vert \psi_0 \rangle$. In the extremely localized limit, $|\alpha|^2+|\beta_0|^2\approx 1$. 
We thus define another quantity to characterize the localization,
\be
W=|\alpha|^2+|\beta_0|^2=|\langle \psi_0 \vert \psi_I  \rangle\vert^2+|\langle \psi_0 \vert { 0} \rangle\vert^2,
\ee
which reaches its maximum when the eigenstate is maximally localized. $|\psi_0\rangle$ then mainly depends on $H_0$, as its couplings to other states have been quenched in this frustrated model. If $\beta_0$ is comparable to $\alpha$, the above simple analytical expressions are no longer accurate. Nevertheless, qualitative results remain unchanged. 
As shown in Fig. 3, when $J_z=J$, $W$ reaches its maximum 0.98,  and $\alpha=0.89$, $\beta_0=0.43$ when $J^{\prime}/J = -0.3$. Correspondingly, $\overline{\xi}$ is minimized for the same parameters. From the above discussions, we see that the overlap between the initial state and the localized eigenstate, $|\psi_0\rangle$, is given by $|\langle \psi_0|\psi_I\rangle|^2=|\alpha|^2$ and thus there exists a time-independent part in $P(t)=|\langle \psi(0)|\psi(t)\rangle|^2$, $|\langle \psi_0|\psi_I\rangle|^4=|\alpha|^4$. As shown in figure 2, the average values of plateaus in $P(t)$ are indeed $|\alpha|^4$. Therefore, the initial state memory is well preserved when $\alpha$ is large.

\begin{figure*}
\includegraphics[scale=0.27]{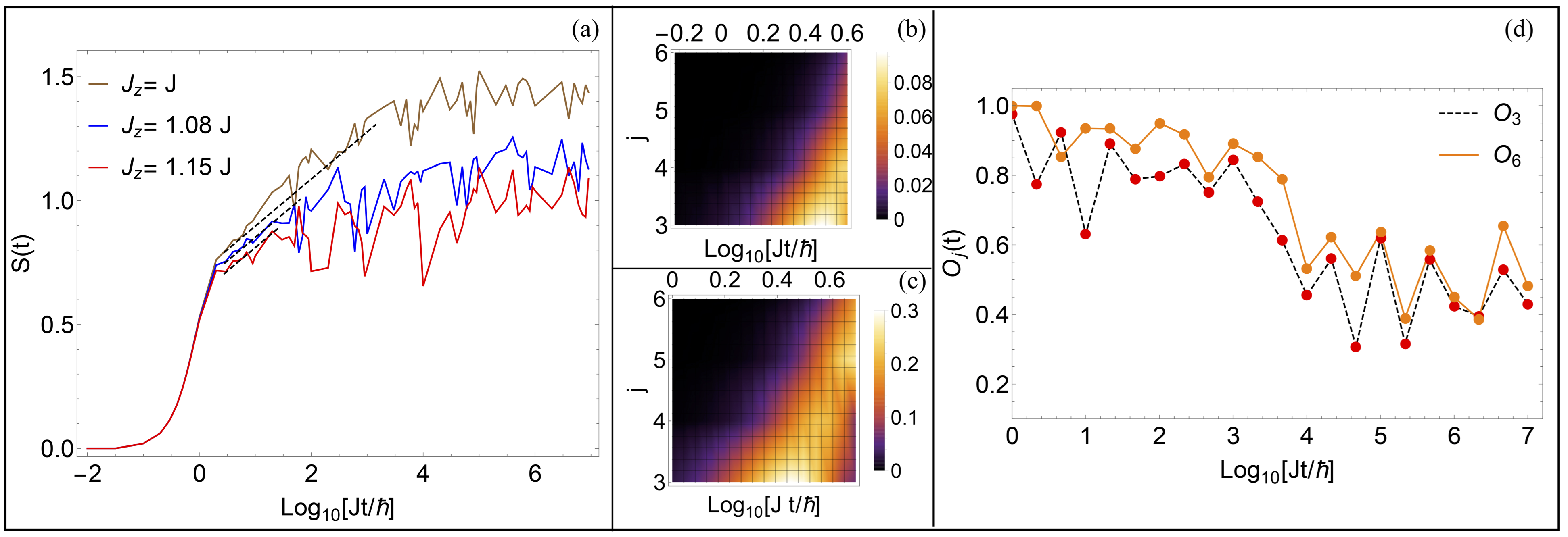}
\caption{
(a) 
After a fast growth at short times, the entanglement entropy grows logarithmically.  
The slopes (dashed lines) are obtained using Eq.(\ref{ent})  with corresponding $\xi$. The same $\lambda=0.19$ has been used for these three curves as these $J'$s are close to each other. 
For this figure $J^{\prime} = -0.25 J$. (b) Density plot of the spin depolarization at site $j$, $1-\sigma^{j}_z$ when $J^{\prime} = -0.25 J$ and $J_z = J$. (c)The dynamics of the Out-of-time-ordered correlator, $O_j$ (eq.(\ref{otocdef}) in the main text). The red dots correspond to $O_3$ and the orange dots correspond to $O_6$. (d) Density plot of $1-O_j$ at short times, where the logarithmic light cone is evident. For this figure, $J^{\prime} = -0.25 J$ and $J_z = J$.
}
\label{f34}
\end{figure*}

{\it Lifetime of quasi-MBL:} In a finite system, there exists an exponentially small coupling between $|\psi_0\rangle$ and $\psi_1\rangle$ such that the eigenstates become $|\psi_{\pm}\rangle=\frac{1}{\sqrt{2}}(|\psi_0\rangle\pm |\psi_1\rangle)$ with corresponding energies $\epsilon_{\pm}$. The exponentially small energy splitting $\Delta \epsilon=\epsilon_+-\epsilon_-\sim e^{-L/\lambda}$, where $\lambda$ is a length scale proportional to the localization length $\xi$, sets up $t^*$,  the lifetime of quasi-MBL, when $P(t)$ first becomes vanishing small. 
In figure 2, we quantify $t^*$ by computing the time when $P(t)$ falls below $0.05$. We plot how $t^*$ varies with the system size. Our results suggest that $t^*$ does grow exponentially, a characteristic feature of quasi-MBL. When $t>t^*$, $|\psi_1\rangle$ becomes important in the quantum dynamics. Whereas we focus on the open boundary condition here, similar results can be obtained for periodic boundary conditions \cite{suppmat}.

{\it Entanglement Entropy and Out-of-time ordered Correlator:} An important signature of MBL and quasi-MBL that distinguishes them from both single particle Anderson localization and the ergodic phase is the dynamics of the half-chain entanglement entropy \cite{BardarsonMBLPRL,BardarsonMBLNJP}, $S=-\rho_L \ln \rho_L$,
where the reduced density matrix of the left(right) half of the chain, $\rho_L = {\rm Tr}_R(\vert\psi\rangle\langle \psi\vert$ is computed by tracing over the degrees of freedom of the right(left) half of the chain. After a fast increase in a very short time, $S$ grows logarithmically in time till the entropy saturates due to the finite size of the system,  unlike single particle Anderson localization, where
the entanglement entropy does not grow at all, or the ergodic phase, where $S$ has a power-law spreading.

As shown in figure \ref{f34}(a), the time dependence of $S$ 
for our initial state shows the characteristic behavior of quasi-MBL. In a short time scale $\sim \hbar/J$, $S$ increases very fast when $|\psi_I\rangle$ mixes with $|0\rangle$. After that, $S$ grows logarithmically slowly due to the localized nature of $|\psi_0\rangle$, the eigenstate of the Hamiltonian having a large overlap with the initial state, $|\psi_I\rangle$. This can be qualitatively understood from tracing the spin depolarization \cite{GarrahanPRB}. The exponential spatial decay of the eigenstate $|\psi_0\rangle$ implies that the spin at a site $j> 0$ ($j<0$) significantly deviates from $1/2$ ($-1/2$) at a time scale $t_d\sim {t_0} e^{-|j|/\xi}$, where $t_0$ set-up by $J$ and $J^{\prime}$ is independent of $j$. Therefore, we define a length scale, $x_t\sim \xi \log (t/t_0)$, measured from the center of the spin chain. At a given time $t$, only the regime, $x\in [-x_t, x_t]$, contributes to $S$, as spins outside this regime are still fully polarized as the initial state.  Thus, $S$ can be written as \cite{SerbynMBLPRL}, 
\be
S\sim \lambda \xi \log (t/t_0),\label{ent}
\ee
where $\xi$ is the width of the tail of $|\psi_0\rangle$ given by Eq.(\ref{loclength}) and $\lambda$ is a numerical factor. As shown in Fig.4(a), we have used the same $\lambda=0.19$ to fit the logarithmic growth of $S$ for a few different sets of parameters. At later times, $S$ begins to saturate when the excitations gradually propagate towards the edge of our system. 

Since it is, in general, a grand challenge to measure entanglement entropy, it was recently suggested that
an appropriate out-of-time ordered (OTO) correlator can trace the propagation of information in MBL
\cite{Zhaieeotoc}. Such correlators have been measured in recent experiments \cite{Reyotoc,Zhaiotoc}. Furthermore, any OTO correlator decays to zero for an ergodic system while certain OTO correlators remain non-zero for very long times in MBL \cite{Fradkinotoc,HeLuotocPRB2017,Chenotoc,SwingleChowdhuryotocPRB2017}. For our model, we compute an OTO correlator of the form:
\be 
{O}_j = \langle \sigma_{\frac{L}{2}}^{z}(0) \sigma_{\frac{L}{2}+j}^{z} (t) \sigma_{\frac{L}{2}}^{z}(0) \sigma_{\frac{L}{2}+j}^{z} (t) \rangle. 
\label{otocdef}
\ee
Numerical results of OTO correlators are shown in Fig. 4(b). A logarithmic light cone is evident at short times. This can also be traced back to the logarithmic dependence of $t_d$, the time scale for the spin depolarization, on the distance to the origin. When the spin at site $j$ remains fully polarized as that in the initial state, $O_j$ is equal to 1. Only when a depolarization occurs, $O$ begins to deviate from 1. At long times, different $O_j$ saturate to the same value, similar to OTO correlators in MBL \cite{Fradkinotoc,HeLuotocPRB2017}.

{\it Manipulating the qubit:} As discussed before, in the localized regime, i.e., $|\alpha|^2+|\beta_0|^2\approx 1$, and the eigenstate $|\psi_0\rangle$ can be well approximated by $|\psi_0\rangle=\alpha|\psi_I\rangle+\beta_0|0\rangle$. Though such a pseudospin-1/2 arises in a many-body system, it distinguishes from other spins embedded in a large environment, where decoherence is, in general, severe. Here, the pseudospin-1/2, $\vec{\tau}$, comprises only two localized states and is largely decoupled from the rest of the system. This is precisely the underlying mechanism for the long life time and other intriguing properties of quasi-MBL in our system. These results suggest that we could further manipulate $\vec{\tau}$ and implement it as a potential qubit. To this end, we consider a generalization of the Hamiltonian in Eq.(\ref{H}), 
\bea
H' &=& \sum_i (Je^{i\theta}S_i^{\dagger}S_{i+1}+J^{\prime}e^{2i\theta} S_i^{\dagger}S_{i+2}+h.c) \nonumber\\
&+& J_z {S}^z_i{S}^z_{i+1}+(V_{L/2} {S}^z_{L/2}+V_{L/2+1} {S}^z_{L/2+1}) ,
\label{H'}
\eea
where $V_{L/2}$ and $+V_{L/2+1}$ are local potentials at the sites $\frac{L}{2}$ and $\frac{L}{2}+1$ respectively. The phases $e^{i\theta}$ and $e^{2i\theta}$ added to $J$ and $J'$, respectively, do not affect the localization properties and any other previously discussed results. Meanwhile, the corresponding local Hamiltonian $H_0'$ can be written as  
 \bea
H_{0}'=\Delta |\psi_I\rangle\langle \psi_I|+(Je^{i\theta} |\psi_I\rangle\langle 0|+h.c.),
\eea
where $\Delta = \frac{J_z}{2}+(V_{i/2}-V_{i/2+1})$. Thus, controlling $\Delta$ and $\theta$ allows one to rotate $\vec{\tau}$ essentially arbitrarily on the Bloch sphere such that it may be used as a potential qubit for quantum information processing. 
  
{\it Experimental realizations:} A variety of quantum emulators can be used to realize spin models that would exhibit quasi-MBL. Several schemes have been proposed to realize frustrated spin models like the one that we study in ion-traps \cite{RichermePRBprop, Solanoscirep} or with atoms trapped in a photonic crystal waveguides \cite{HungPNAS2016}.  Alternatively, one could make use of the well established mapping between spin-1/2 particles and hard core bosons \cite{SachdevBook}. Our frustrated spin model thus maps to a boson model, in which $J$ and $J'$ correspond to the nearest and the next nearest neighbor tunneling, and $J_z$ is the nearest neighbor interaction. In such a bosonic model, kinetic frustration can be engineered in set-ups involving cold atoms loaded in driven optical lattices \cite{BlochHarper1,BlochHarper2,KetterleHarper1,KetterleHarper2}. 
Tilting the optical lattice by an external field, the bare tunneling is suppressed and Raman lasers induce photo-assisted tunneling, the phase of which could be tuned. Whereas current experiments have realized Raman dressed nearest neighbor tunneling, the same technique can be directly implemented to produce a next nearest neighbor tunneling to access a frustrated system. Interestingly, such frustrated model has been obtained using a momentum space lattice \cite{Gadwayfrust}. It is promising that our results will be relevant to experiments in the near future. 
 
{\it Summary and Outlook:} 
We have shown the existence of quasi-MBL in a frustrated 1D spin chain without disorder. 
Localized many-body eigenstates lead to memory retention for exponentially long times, logarithmic growth of entanglement entropy and a logarithmic light cone of an out-of-time-ordered correlator.  
Furthermore, we discuss how the localized eigenstates can be potentially manipulated as a qubit.
Our work shows that the interplay of frustration and interaction can give rise to quasi-MBL in a broad class of disorder free systems. 

{\it Acknowledgements:} This work is supported by startup funds from Purdue University.


\begin{thebibliography}{10}
\bibitem{Anderson1958}P. W. Anderson, {Phys. Rev.} {\bf 109}, 1492 (1958).
\bibitem{BLL2006}  D. M. Basko, I. L. Aleiner, and B. L. Altshuler, Ann. Phys. (Amsterdam) {\bf 321}, 1126 (2006).
\bibitem{PalHuse2010} A. Pal and D. A. Huse, Phys. Rev. B 82, 174411 (2010).
\bibitem{MoorePRL2012} J. H. Bardarson, F. Pollmann, and J. E. Moore, Phys. Rev.Lett. {\bf 109}, 017202 (2012).
\bibitem{HusePRB2014MBL} D. A. Huse, R. Nandkishore, and V. Oganesyan, Phys. Rev.
B {\bf 90}, 174202 (2014).
\bibitem{KhemaniMBLPRX} V. Khemani, S. P. Lim, D. N. Sheng, D. A. Huse, Phys. Rev. X {\bf 7}, 021013 (2017).
\bibitem{VoskMBLPRX} R. Vosk, D. A. Huse, and E. Altman, Phys. Rev. X {\bf 5}, 031032 (2015).
\bibitem{SerbynPRLMBL} M. Serbyn, Z. Papi\'{c}, and D. A. Abanin, Phys. Rev. Lett.
{\bf 111}, 127201 (2013). 
\bibitem{AltlandPRLMBL2017} A. Altland and T. Micklitz, Phys. Rev. Lett. {\bf 118}, 127202 (2017)
\bibitem{HuseMBLPRB2013} D. A. Huse, R. Nandkishore, V. Oganesyan, A. Pal, and
S. L. Sondhi, Phys. Rev. B {\bf 88}, 014206 (2013).
\bibitem{NayakTC} D. V. Else, B. Bauer, and C. Nayak, Phys. Rev. Lett. {\bf 117}, 090402 (2016)
\bibitem{YaoTC} N. Y. Yao, A. C. Potter, I.-D. Potirniche, A. Vishwanath, Phys. Rev. Lett. {\bf 118}, 030401 (2017).
\bibitem{BahriNatComm} Y. Bahri, R. Vosk, E. Altman, and A. Vishwanath, Nat. Commun. {\bf 6}, {7341} (2015).
\bibitem{YaoQCMBL1} S. Choi, N. Y. Yao, S. Gopalakrishnan, and M. D. Lukin,
ArXiv e-prints (2015), arXiv:1508.06992 [quant-ph].
\bibitem{YaoQCMBL2} N. Y. Yao, C. R. Laumann, and A. Vishwanath, ArXiv
e-prints (2015), arXiv:1508.06995 [quant-ph].
\bibitem{NandkishoreMBLrev} R. Nandkishore and D. A. Huse, Annu. Rev. Condens. Matter Phys. {\bf 6}, 15 (2015) and references therein.
\bibitem{BlochMBL1} M. Schreiber, S. S. Hodgman, S. Bordia, H. P. Lüschen, M. H. Fischer, R. Vosk, E. Altman, U. Schneider, and I. Bloch, Science {\bf 349}, 842 (2015).
\bibitem{BlochMBL2}  J.-Y. Choi, S. Hild, J. Zeiher, P. Schauss, A. Rubio-Abadal,
T. Yefsah, V. Khemani, D. A. Huse, I. Bloch, and C. Gross, Science {\bf 352}, 1547 (2016).
\bibitem{MonroeMBL}  J. Smith, A. Lee, P. Richerme, B. Neyenhuis, P. W. Hess, P. Hauke, M. Heyl, D. A. Huse, and C. Monroe,
Nature Physics {\bf 12}, 907 (2016).
\bibitem{MonroeTC} J. Zhang, P. W. Hess, A. Kyprianidis, P. Becker,
A. Lee, J. Smith, G. Pagano, I.-D. Potirniche, A. C.
Potter, A. Vishwanath, N. Y. Yao, and C. Monroe,
Nature {\bf 543}, 217 (2017).
\bibitem{YaocleanMBL} N. Y. Yao, C. R. Laumann, J. I. Cirac, M. D. Lukin, and J. E.
Moore, Phys. Rev. Lett. {\bf 117}, 240601 (2016).
\bibitem{GrovercleanMBL}  T. Grover and M. P. A. Fisher, J. Stat. Mech. (2014) P10010.
\bibitem{MullercleanMBL} M. Schiulaz, A. Silva, and M. Muller, Phys. Rev. B {\bf 91}, 184202 (2015)
\bibitem{AbanincleanMBL} Z. Papic, E. M. Stoudenmire, and D. A. Abanin, Ann. Phys.
(N. Y). {\bf 362}, 714 (2015).
\bibitem{HuveneerscleanMBL} W. De Roeck and F. Huveneers, Phys. Rev. B {\bf 90}, 165137 (2014).
\bibitem{GarrahanJSM} J. M. Hickey, S. Genway, and J. P. Garrahan, J. Stat. Mech.:
Theor. Exp. (2016) 054047.
\bibitem{GarrahanPRB} ] M. van Horssen, E. Levi, and J. P. Garrahan, Phys. Rev. B {\bf 92}, 100305 (2015).
\bibitem{SmithPRLMBL} A. Smith, J. Knolle, D.L. Kovrizhin, and R. Moessner, Phys. Rev. Lett. {\bf 118}, {266601} (2017).
\bibitem{ScardicchioPRLMBL2018} M. Brenes, M. Dalmonte, M. Heyl, and A. Scardicchio, Phys. Rev. Lett. {\bf 120}, 030601 (2018).
\bibitem{Arconnatcomm} M. Pregelj, A. Zorko, O. Zaharko, H. Nojiri, H. Berger, L. C. Chapon, and D. Arcon,  Nat. Commun. {\bf 6}, 7255 (2015).
\bibitem{LauchliPRB2009} J. Sudan, A. L\"{u}scher, and A. M. L\"{a}uchli,  Phys. Rev. B {\bf 80}, 140402(R) (2009).
\bibitem{FurusakiPRB2008} T. Hikihara, L. Kecke, T. Momoi, and A. Furusaki, Phys. Rev. B {\bf 81}, 224433 (2010)..
\bibitem{SerbynMooreMBL} M. Serbyn and J. E. Moore, Phys. Rev. B {\bf 93}, 041424 (2016).
\bibitem{suppmat} See Supplementary Materials for detailed information.
\bibitem{BardarsonMBLPRL} J. H. Bardarson, F. Pollmann, and J. E. Moore Phys. Rev. Lett. {\bf 109} 017202 (2012).
\bibitem{BardarsonMBLNJP} R. Singh, J. H. Bardarson, and F. Pollmann, New J. Phys. {\bf 18}, 023046 (2016).
\bibitem{Zhaieeotoc} R. Fan, P. Zhang, H. Shen, and H. Zhai, Sci. Bull. {\bf 62} (10), 707-711 (2017).
\bibitem{Reyotoc} M. G\"{a}rttner, J. G. Bohnet, A. Safavi-Naini, M. L. Wall, J. J. Bollinger, and A. M. Rey, \NATP{13}{781–786}{2017}.
\bibitem{Zhaiotoc} J. Li, R. Fan, H. Wang, B. Ye, B. Zeng, H. Zhai, X. Peng, and J. Du, \PRX{7}{031011}{2017}.
\bibitem{Fradkinotoc} X. Chen, T. Zhou, D. A. Huse, and E. Fradkin, Annalen der Physik {\bf 529}, 1600332 (2017).
\bibitem{HeLuotocPRB2017} R.-Q. He and Z.-Y Lu, \PRB{95}{054201}{2017}.
\bibitem{Chenotoc} Y. Huang, Y.-L. Zhang,and  X. Chen, Annalen der Physik {\bf 529}, 1600318 (2017).
\bibitem{SwingleChowdhuryotocPRB2017} B. Swingle and D. Chowdhury,  Phys. Rev. B {\bf 95}, 060201(R) (2017). 
\bibitem{SerbynMBLPRL} M. Serbyn, Z. Papi\'{c}, and D. A. Abanin, Phys. Rev. Lett. {\bf 111}, 127201 (2013).
\bibitem{HuseMBLPRB2014} D. A. Huse and V. Oganesyan, Phys. Rev. B {\bf 90}, 174202 (2014).
\bibitem{RichermePRBprop} A. Bermudez, L. Tagliacozzo, G. Sierra, and P. Richerme 
Phys. Rev. B. {\bf 95}, 024431 (2017)
\bibitem{Solanoscirep} I. Arrazola, J. S. Pedernales, L. Lamata, and E. Solano, Sci. Rep. {\bf 6}, 30534 (2016).
\bibitem{HungPNAS2016} C.-L. Hung, A. González-Tudela, J. I. Cirac, and H. J. Kimble, Proc. Natl. Acad. Sci. USA {\bf 113}, E4946–E4955 (2016).
\bibitem{SachdevBook} S. Sachdev, {\it Quantum Phase Transitions} (Cambridge University Press, 2001).
\bibitem{BlochHarper1} M. Aidelsburger, N. Atala, M. Lohse, J. T. Barreiro, B. Paredes, and I. Bloch, Phys. Rev. Lett. {\bf 111}, 185301 (2013).
\bibitem{BlochHarper2} M. Aidelsburger, M.Lohse, C. Schweizer, M. Atala, J. T. Barreiro, S. Nascimbene, N. R. Cooper, I. Bloch, and N. Goldman, Nat. Phys. {\bf 11}, 162 (2014).
\bibitem{KetterleHarper1} H. Miyake, G. A. Siviloglou, C. J. Kennedy, W. C. Burton, and W. Ketterle, Phys. Rev. Lett. {\bf 111}, 185302 (2013).
\bibitem{KetterleHarper2} C. J. Kennedy, W. C. Burton, W. C. Chung, and W. Ketterle, Nat. Phys. {\bf 11}, 859 (2015).
\bibitem{Gadwayfrust} F. A. An, E. J. Meier, J. Ang’ong’a, and B. Gadway, Phys. Rev. Lett. {\bf 120}, 040407 (2018).
\end{thebibliography}
\end{document}

\title{{Supplementary Material: Frustration induced quasi-many-body localization without disorder}}
\author{Sayan Choudhury,$^{1}$ Eun-Ah Kim,$^{2}$ and Qi Zhou$^{1}$}
\email{zhou753@purdue.edu}
\affiliation{$^{1}$Department of Physics and Astronomy, Purdue University, 525 Northwestern Avenue, West Lafayette, IN 47907, USA\\
$^{2}$Department of Physics, Cornell University, Ithaca, New York 14853, USA}
\maketitle
\section{Many-body localization with periodic boundary conditions}
\begin{figure}
\scalebox{0.48}{\includegraphics{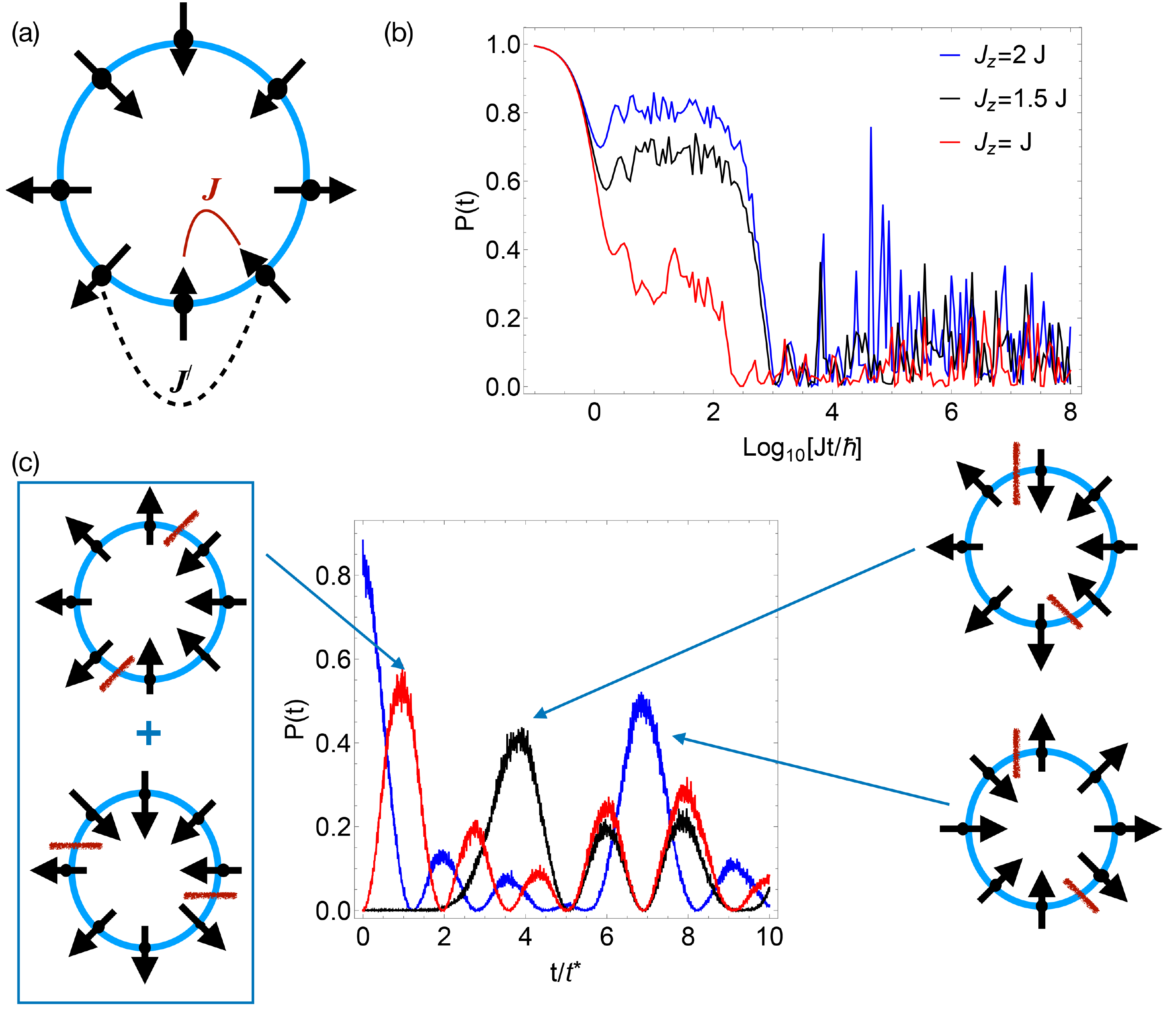}}
\caption{(a) Schematic of the spin-chain with periodic boundary conditions. (b) Memory retention of the initial state for a 12 site periodic spin-chain: When the system initially has only two domain walls, then for a wide parameter regime the memory of the initial state (as characterized by $P(t)$ as defined in eq.(2) of the main text) is retained for a long time. For this figure, we have fixed $J^{\prime} = -0.25 J$. (c) When the memory of the initial conditions is lost ($t \sim t^*$), other states with two domain walls appear in the dynamics. The blue line shows $P(t)$ for an initial state with two domain walls. The red line shows the weight of states where the domain walls have moved by one site either clockwise or counter-clockwise. The black curve shows the weight of states where both the domain walls have moved by L/2 sites. In the schematic representation of these states, arrows pointing outward(inward) represent spin-up(spin-down) and the red(thick) lines represent domain walls.}
\label{fs1}
\end{figure}

In the main text, we studied the dynamics of a frustrated one-dimensional spin chain with open boundary conditions. We found that the chain exhibited signatures of quasi-many body localization(MBL) when it is initially prepared in a state with only one domain wall. In this section we study the dynamics of the same chain with periodic boundary conditions. A schematic of the model is shown in figure \ref{fs1}(a). The Hamiltonian for this model is given by:
\bea
H = \sum_{i<L} \left( \frac{J_z}{2} {S}^z_i{S}^z_{i+1}+\frac{J}{2} S_i^{\dagger}S_{i+1}+\frac{J^{\prime}}{2} S_i^{\dagger}S_{i+2}\right) + \left(\frac{J_z}{2} {S}^z_L{S}^z_{1} + \frac{J}{2} S_L^{\dagger}S_{1}+\frac{J^{\prime}}{2} S_{L-1}^{\dagger}S_{1}+ S_{L}^{\dagger}S_{2}\right)+h.c, \label{H}
\eea
We find that the periodic chain also exhibits signatures of quasi-MBL when it is prepared in initial state with two domain walls. This initial state is an analog of the state with one domain wall in the open chain, $\vert \psi_I \rangle$. Just as in the main text, we characterize the memory retention of the initial state by:
\be
P(t) = \vert \langle \psi(t=0) \vert \psi(t)\rangle \vert^2
\ee
As shown in figure \ref{fs1}(b), the spin chain retains the memory of the initial state for very long times. We can gain further insight by examining which states are occupied when the memory of the initial state is lost. We find that when $t> t^*$ ($t^*$ being the lifetime of the quasi-MBL), other states with two domain walls appear in the dynamics. In particular, when $t \sim t^*$, states where both domain walls have moved by one site (clockwise or counter-clockwise) contribute a large weight to the many-body wavefunction. Our results are shown in figure \ref{fs1}(c).

\section{Analytic expression for the Localized eigenstate}

In this section, we provide an explicit analytical expression for the localized eigenstate, $\vert \psi_0 \rangle$. As shown in eq.(3) in the main text, the localized eigenstate, $\vert \psi_0 \rangle$ can be written as:
\be
\vert\psi_0 \rangle =
\alpha|\psi_I\rangle+\beta_0  \vert { 0}\rangle+\sum_{j\neq 0} \beta_j \vert { j}\rangle+\ldots, \label{psi0}
\ee
where $|\psi_I\rangle$ has one domain wall, while $\vert { 0}\rangle$ and $\vert { j}\rangle$ has three domain walls, 
and $\ldots$ represents terms with negligible weights. 
In the main text, we have defined an effective Hamiltonian whose Hilbert space only comprises states with one or three domain walls. As shown in Fig. 2, this effective Hamiltonian can capture both $P(t)$ and the localized eigenstate very well. For this effective Hamiltonian, we can derive an analytical expression for the localized eigenstate, $\vert \overline{\psi}_0 \rangle$ which has the form:
\be
\vert\overline{\psi}_0 \rangle = \alpha|\psi_I\rangle+\beta_0  \vert { 0}\rangle+\sum_{j\neq 0} \beta_j \vert { j}\rangle, \label{psibar0}
\ee
Using the Schr\"{o}dinger equation, any $\beta_j$ can be written as:
\be
\beta_j = \frac{J}{\Delta} \beta_{j-1} + \frac{J^{\prime}}{\Delta} \beta_{j-1},
\label{rec}
\ee
where $\Delta \approx 2 J_z$. When $l \ge 1$, we solve eq.(\ref{rec}) to obtain an expression for $\beta_l$:
\bea
\beta_{2 l-1} &=& \frac{J^{\prime}}{\Delta} P1_{l-1} \alpha + P2_{l-1} \beta_0, \\
\beta_{2 l} &=& \frac{J^{\prime}}{\Delta} P2_{l-1} \alpha + P1_{l} \beta_0,
\eea
where
\bea
P1_l &=& \frac{2^{-l}}{(g_1 g_0)^2}\left(\frac{J^4}{\Delta^4} g_1^{2l}+ \frac{4 J^{\prime 2}}{\Delta^2}(g_1^{2l}+ g_2^{2l})+ \frac{J^2}{\Delta^2}(\frac{J}{\Delta} g_0 g_1^{2l} +\frac{J^{\prime}}{\Delta}(g_2^{2l}+5 g_1^{2 l})) + \frac{J^{\prime}J}{\Delta^2} g_0(g_2^{2l} +3 g_1^{2l})\right),\\
P2_l &=& \frac{\Delta}{J}\frac{ 2^{-l}}{ (g_1 g_0)^2} \left((\frac{J^6}{\Delta^6}+\frac{J^5}{\Delta^5} g_0)g_1^{2 l})+ (\frac{6 J^4 J^{\prime}}{\Delta^5}+\frac{4J^3J^{\prime}}{\Delta^4} g_0)g_1^{2 l}) + \frac{J^{\prime 2}J}{\Delta^3}(8\frac{J}{\Delta}g_1^{2l}-2 g_0(g_2^{2l}-g_1^{2l})\right),
\eea
and 
\bea
g_0 &=& \sqrt{\left(\frac{J^2}{\Delta ^2}+\frac{4 J^{\prime}}{\Delta }\right)}, \\
g_1 &=& \sqrt{\frac{J^2}{\Delta ^2}+\frac{2 J^{\prime}}{\Delta }+ \frac{J}{\Delta} g_0},\\
g_2 &=& \sqrt{\frac{J^2}{\Delta ^2}+\frac{2 J^{\prime}}{\Delta }- \frac{J}{\Delta} g_0}.
\eea
\begin{figure*}
\subfigure[$\, $]{\raisebox{0.35cm}{\scalebox{0.345}{\includegraphics{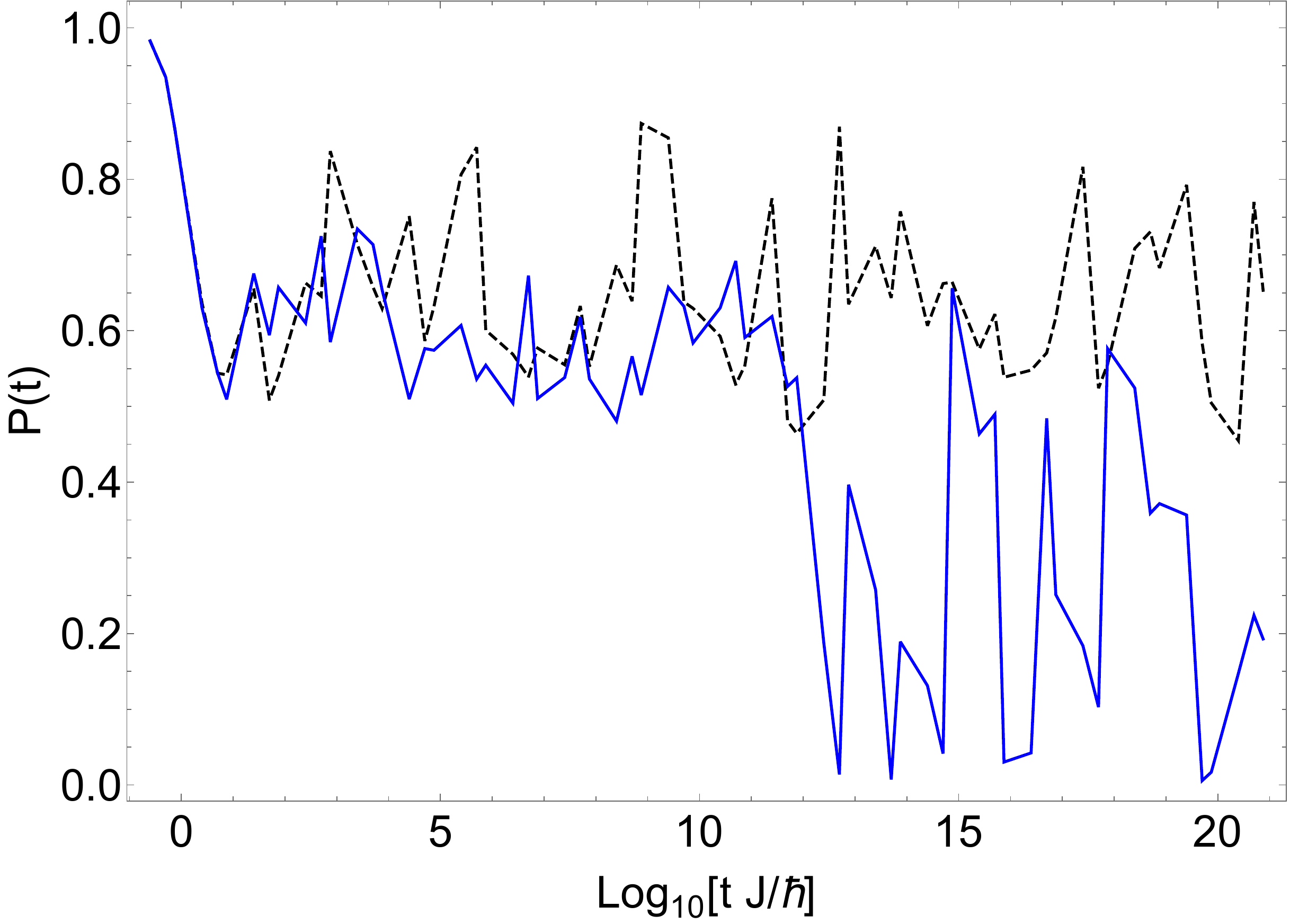}}}}
\,\,\,\,\subfigure[$\,$]{\raisebox{0.35cm}{\scalebox{0.342}{\includegraphics{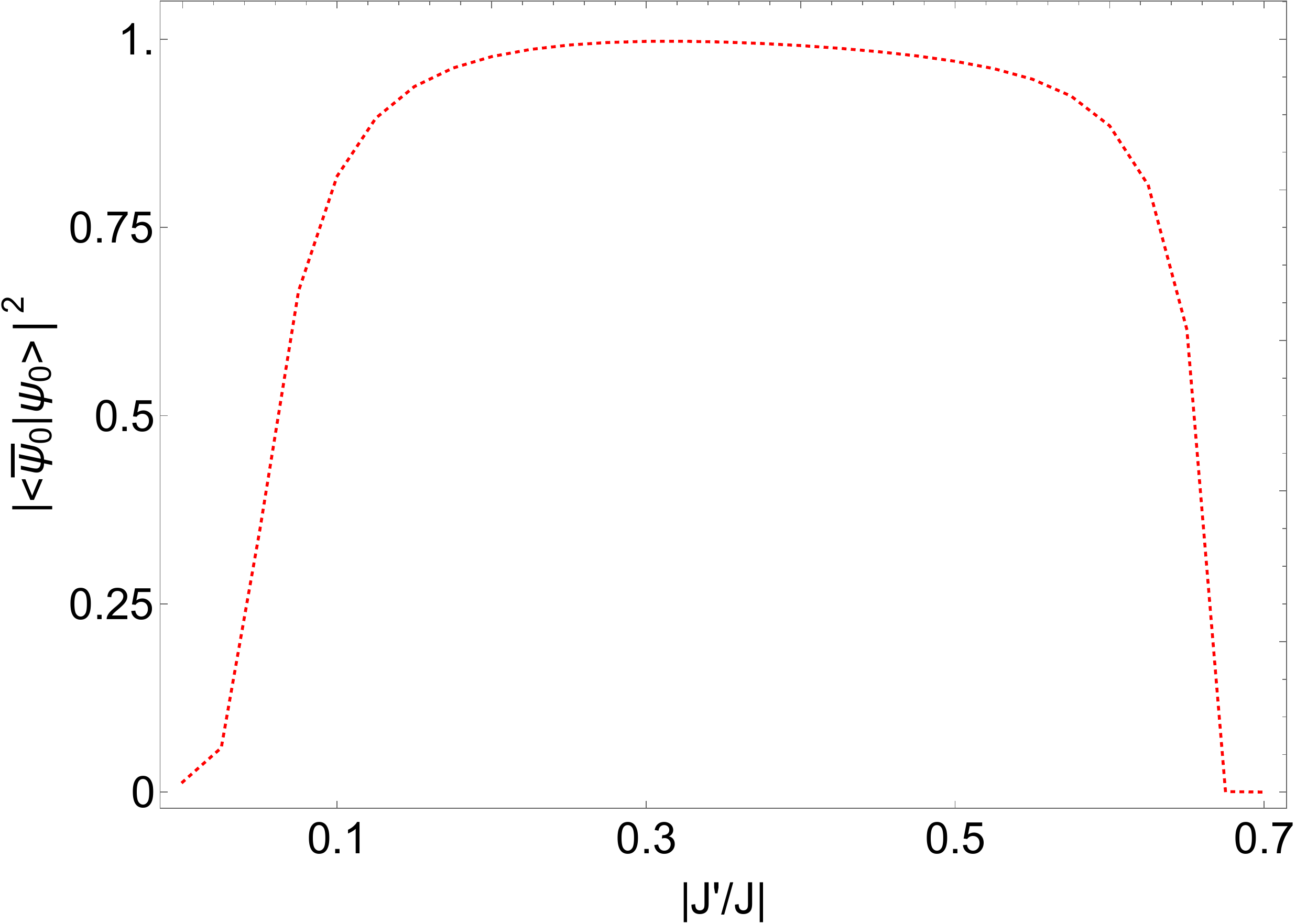}}}}
\caption{(a) Memory retention of the initial state with only one domain wall. For this figure, $J^{\prime} = -0.3 J$ and $J_z = J$. The blue(thick) line shows the exact dynamics for a 14 site model, while the black(dashed) line shows the dynamics in the effective model 
$H_{\text{eff}}$, where the Hilbert space comprises states with only 1 or 3 domain walls. The effective model can capture the dynamics of the initial state upto long times (b) Overlap of the localized eigenstate in the effective model, $\vert \overline{\psi}_0 \rangle$ with the localized eigenstate in the full model, $\vert\psi_0 \rangle$. For a wide parameter range, this overlap is very close to 1.}
\label{fs2}
\end{figure*}

These lengthy equations simplify in the limit when $(\frac{J}{\Delta})^2 \ge 4 J^{\prime}/\Delta$. In that limit, we obtain:
\bea
P1_l &=& 2^{-l} \frac{g_1^{2 l+1} - g_2^{2l+1}}{g_1-g_2},\\
P2_l &=& 2^{-(l+1/2)} \frac{g_1^{2 l+2} - g_2^{2l+2}}{g_1-g_2}.\\
\eea
Furthermore, when $\beta_0$ is well approximated by $J/\Delta\alpha$, we obtain:
\bea
\beta_{2 l-1} &=& P1_l \alpha, \\
\beta_{2l} &=& P2_l \alpha.
\eea
This implies the following simple form for $\beta_j$:
\bea
\beta_{j} &=& \frac{1}{2^{(j+1)/2}} \frac{g_1^{j+2} - g_2^{j+2}}{g_1-g_2} \alpha \nonumber\\
&=& \left(A \exp(-j/\xi_1) + B \exp(-j/\xi_2)\right) \alpha.
\eea
This is exactly the form of $\beta_j$ given in eq.(8) of the main text.